\documentclass[a4paper,11pt]{article}
\pdfoutput=1 

\usepackage{jcappub} 

\usepackage[T1]{fontenc} 
\usepackage{breqn}
 \def\be   {\begin{equation}}   \def\ee   {\end{equation}}
 \def\ba   {\begin{array}}      \def\ea   {\end{array}}
 \def\bea  {\begin{eqnarray}}   \def\eea  {\end{eqnarray}}
 \def\bean {\begin{eqnarray*}}  \def\eean {\end{eqnarray*}}
\definecolor{verde}{rgb}{0,0.5,0}

\title{\boldmath 
Gravitational Wave non-Gaussianity from  trans-Planckian Quantum Noise

}

\usepackage{appendix}

\author[a,b]{Mattia Cielo,}
\author[c,d]{Matteo Fasiello,}
\author[a,b]{Gianpiero Mangano,}
\author[a,b]{Ofelia Pisanti}

\affiliation[b]{Dipartimento di Fisica ``Ettore Pancini”, Università degli studi di Napoli ``Federico II”, Complesso Univ. Monte S. Angelo, I-80126 Napoli, Italy}
\affiliation[a]{INFN - Sezione di Napoli, Complesso Univ. Monte S. Angelo, I-80126 Napoli, Italy}
\affiliation[c]{Instituto de Física Téorica UAM/CSIC, calle Nicolás Cabrera 13-15, Cantoblanco, 28049,
Madrid, Spain}
\affiliation[d]{Institute of Cosmology $\&$ Gravitation, University of Portsmouth, PO1 3FX, UK}

\emailAdd{mattia.cielo@na.infn.it}
\emailAdd{matteo.fasiello@csic.es}
\emailAdd{gmangano@na.infn.it}
\emailAdd{pisanti@na.infn.it}

\abstract{
We examine the effect of a trans-Planckian phase on the dynamics of inflationary tensor perturbations. To remedy the fact that this regime is not fully captured by standard perturbation theory, we introduce an effective quantum noise source, whose role is regulated by the energy scale $\Lambda$. The presence of the source modifies the initial conditions for the tensor modes, leaving a distinct imprint. We study the amplitude and shape of the gravitational wave bispectrum of the model and compare these with their counterparts obtained under the assumptions of Bunch-Davies initial conditions and $\alpha$-vacua states.
Depending on the value of the scale $\Lambda$, we find distinctive signatures  associated with both the bispectrum shape and the non-linear parameter $f_{\rm NL}$.}

\begin{document}
\maketitle
\flushbottom

\section{Introduction}

Inflation stands as one of the main pillars of our current understanding of cosmology. It solves the puzzles associated with the Hot Big Bang model and provides a mechanism to seed the growth of structure \cite{Starobinsky:1980te, Sato:1981ds, PhysRevD.23.347, Linde:1981mu}. These successes of the inflationary paradigm notwithstanding, there is plenty we do not yet know about inflation. Most notably, a host of wildly different inflationary models are compatible with current observations, CMB experiments being those providing the most stringent constraints to date.
The key questions that remain unanswered are the energy scales involved during inflation and the field content of the inflationary Lagrangian. In current parlance, we do not yet know at what energy the ``cosmological collider'' operates nor do we have crucial information on the (self-)interactions characterising the inflationary particle zoo. 

Remarkably, an unprecedented array of cosmological probes is scheduled to become operational in the next decade: it promises to transform the current picture of the early universe and address some of the most pressing issues. What is perhaps most exciting is the prospect of testing more than 20 decades in the frequency of the gravitational wave (GW) signal in the coming years. 
At CMB scales we will soon be able to test the energy scale of inflation by ever-more-stringent bounds on the tensor to scalar ratio $r$ ($\sigma_r\sim 10^{-2}$ for e.g. CMB-S3 experiments \cite{BICEP2:2018kqh, BICEP:2021xfz} and $\sigma_r\sim 10^{-3}$ for CMB-S4 \cite{CMB-S4:2016ple, abazajian2019cmbs4} and LiteBIRD \cite{LiteBIRD:2022cnt}). Large-scale structure experiments such as Euclid and the Rubin Observatory will constrain inflationary interactions down do $\sigma_{f_{\rm NL}}\footnote{ For the definition of $f_{\rm NL}$ see Eq. (\ref{fNLstandard}) for example.} \sim ({\rm a\,few})$, whilst 21cm cosmology promises even more down the line \cite{Munoz:2015eqa}. Given that primordial gravitational waves are a \textit{universal} prediction of inflation, one ought to mention the possibilities afforded by pulsar timing arrays \cite{Campeti:2020xwn, NANOGrav:2023gor, NANOGrav:2023hde, NANOGrav:2023hvm, EPTA:2021crs, EPTA:2023fyk, EPTA:2023sfo, EPTA:2023xxk, Zic:2023gta, Reardon:2023gzh, Reardon:2023zen}, and ground and space-based interferometers \cite{LIGOScientific:2016jlg, Campeti:2020xwn, Caprini:2018mtu, Achucarro:2022qrl}. 

It is worthwhile to also widen the overall perspective and recall that, although the structure we observe today may originate from magnified quantum fluctuations, current data is compatible with both a classical and quantum origin of the patterns we see imprinted in our observables. Intriguingly, key insights on the quantum \textit{vs} classical origin may be inferred from the structure of the cosmological correlation functions \cite{Green:2020whw,Micheli:2022tld}.
Cosmological correlators are also sensitive to the choice of the vacuum or initial state for inflation \cite{Holman:2007na}.  In this work, we will go beyond the standard Bunch-Davies choice for the vacuum and explore other well-motivated initial conditions, such as those known as $\alpha$-vacua \cite{Allen:1985ux, Akama:2020jko, Xue:2008mk, Maldacena:2011nz, Maldacena:2002vr, Kanno:2019gqw, Kundu:2011sg, Ghosh:2022cny, Armendariz-Picon:2006saa, Bahrami:2013isa, Naskar:2019shl, Holman:2007na}. Inflation from non-adiabatic initial states comes with specific signatures, including at the level of gravitational wave non-Gaussianities \cite{Kanno:2019gqw, Kanno:2022mkx}. With respect to existing literature, our work here explores the effect of a non-zero anisotropic stress tensor sourcing gravitational wave production. Given that the inflationary energy regime can be perilously close to the Planck scale, we will also be concerned with the trans-Planckian problem (see e.g. \cite{Alberghi:2003am, Brandenberger:2012aj, Tanaka:2000jw}) and the fact that the standard description of inflation may break down at such energies. We shall tackle the effect of trans-Planckian physics on GWs by adopting an open quantum system description where tensor perturbations effectively interact with their environment. We will model the effect of the environment by means of an anisotropic stress tensor which accounts for the dynamics above an energy threshold $\Lambda$ \cite{Cielo:2022vmo}. 
Some of us argued in \cite{Cielo:2022vmo} that such a source is compatible with a nearly scale-invariant GW spectrum at CMB scales, in full agreement with the latest observations \cite{Planck:2019kim}. In this work, we will investigate the gravitational wave non-Gaussianity stemming from such a setup.

We will add one by one the various layers that make up our system. We start with a simple scalar field minimally coupled to gravity and briefly review the corresponding tensor non-Gaussianity. The next step consists of considering non-adiabatic initial states, specifically $\alpha$-vacua. Last, we consider the effect of non-zero anisotropic stress whose dynamics are regulated by the dimensional quantity  $\Lambda$. \vspace{0.25cm}

This paper is structured as follows. In \textit{Section} \ref{BDv}, we provide a  review of the dynamics associated with the tensor modes during the inflationary period. Additionally, we present a thorough analysis of the essential methodologies required for investigating the bispectrum. Furthermore, we present the outcomes obtained specifically for the Bunch-Davies vacuum selection.
In \textit{Section} \ref{alphav}, we introduce the concept of degeneracy in the vacuum state and highlight the key distinctions from the previous section by incorporating a cut-off energy scale for the vacuum selection. Lastly, in \textit{Section} \ref{source}, we extend the scope of our findings by incorporating a trans-Planckian source that impacts the shape of the bispectrum leaving distinct imprints.

\section{Tensor Bispectrum: with and without a cut-off energy scale}

\subsection{Bunch-Davies vacuum}
\label{BDv}
In this section, we will briefly review the results on tensor non-Gaussianity in the case of the standard Bunch Davies vacuum. Naturally, the starting point for the tree-level power spectrum is the quadratic action describing the behaviour of the tensor degrees of freedom of general relativity, whilst the cubic action is necessary to arrive at the three-point function or, equivalently, the bispectrum. Let us consider as the inflationary Lagrangian that of single-field slow-roll, i.e. general relativity minimally coupled to a ``matter'' Lagrangian $\mathcal{L}_{\rm matter}$:

\bea 
\mathcal{L} = \mathcal{L}_{\rm EH} + \mathcal{L}_{matter} =\sqrt{-g}\Bigg[\frac{1}{2} R + \frac{1}{2} g^{\mu \nu} \partial_{\mu} \partial_{\nu} \phi - V(\phi)\Bigg]\; ,
\label{action1}
\eea
 where we set $M_{\rm Pl}$ to one. Expanding Eq.~(\ref{action1}) around the FLRW solution up to the second order, focusing on standard tensor fluctuations, one obtains the following linearized equations of motion: 

\begin{equation}
    h''_{\bf k} + 2 \mathcal{H} h'_{\bf k} + k^2 h_{\bf k} = 16 \pi G a^2 \Pi_{\bf k}(\tau)
\label{omogenea}
\end{equation}
where the anisotropic stress term $\Pi$ in the RHS is the source term, easily derived from Eq.(\ref{action1}). Here and in \textit{Section} \ref{alphav} we shall work within the context of single-field slow-roll inflation and will therefore neglect the subleading correction due to the $\Pi$ term. By virtue of using the re-scaled field $v \equiv a(\tau) h$ in Eq.~(\ref{omogenea}), one arrives at the  Mukhanov-Sasaki equation: 
\begin{equation}
    v''_{\textbf{k}} + \Big( k^2 - \frac{a''}{a} \Big) v_{\textbf{k}} = 0\; ,
\end{equation}
which admits as a solution the following combination of the Bessel functions: 
\begin{equation}
    v_k(\tau) = A_{k}Y_\nu(k, \tau) + B_{k}J_\nu (k, \tau)\; ,
\end{equation}
where $\nu \approx 3/2$ for a quasi de Sitter background (up to the first slow-roll corrections). 
\\

\begin{figure}
    \centering
    \includegraphics[scale=0.49]{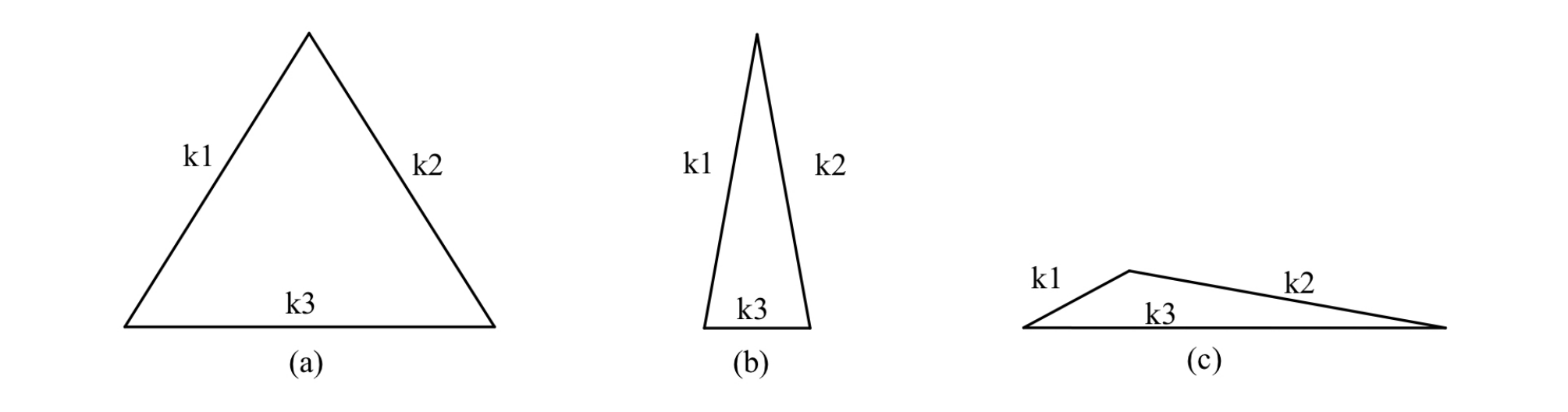}
    \caption{Representation of the three configurations for the momenta $(k_1, k_2, k_3)$. Case (a): \textit{Equilateral} configuration meaning that $k_1 \approx k_2 \approx k_3$. Case (b): \textit{Squeezed} configuration, $k_3 \ll k_1 \approx k_2$ . Case (c): \textit{Folded} shape, $k_3 \approx k_1 + k_2$.}
    \label{fig:triangles}
\end{figure}

In the following, we will employ the in-in formalism to tackle second and higher-order correlation functions. The tree-level contribution to the tensor three-point function reads:
\begin{equation}
    _{in}\langle 0| \hat{h}^{s_1}_{\mathbf{k}_1}(\tau) \hat{h}^{s_2}_{\mathbf{k}_2}(\tau) \hat{h}^{s_3}_{\mathbf{k}_3}(\tau) |0\rangle_{in} = -i \int^\tau_{-\infty} d \tau' a(\tau') \langle \Big[ \hat{h}^{s_1}_{\mathbf{k}_1}(\tau) \hat{h}^{s_2}_{\mathbf{k}_2}(\tau) \hat{h}^{s_3}_{\mathbf{k}_3}(\tau), H^{(3)}_I(\tau')\Big]  \rangle \; ,
    \label{ininformula}
\end{equation}
where the operator $\hat{h}$ is such that $ \hat{h}_{\textbf{k}} = h_{\textbf{k}} \hat{a}_{\textbf{k}} + h^*_{-\textbf{k}} \hat{a}^{\dag}_{-\textbf{k}}  $. 
The gravitational self-interactions of general relativity making up the cubic Hamiltonian in the interaction picture are given by \cite{Chen:2010xka, Wang:2013zva, Maldacena:2002vr, Dimastrogiovanni:2022afr}: 
\begin{equation}
    H^{(3)}_{I}(\tau) = - \frac{1}{M^4_{Pl}}\int d^3x\, a(\tau) \Big( h_{ik}h_{jl} - \frac{1}{2}h_{ij}h_{kl} \Big) \partial_k \partial_l h_{ij} \; .
    \label{Hint}
\end{equation}
Following standard practice (see \cite{Gao:2011vs, Akita:2015mho, Naskar:2019shl}), we write the spectrum as:

\begin{equation}
\langle h_{ij}(\textbf{k}_1) h_{kl}(\textbf{k}_2)\rangle = (2\pi)^3 \delta^{(3)}(\textbf{k}_1+ \textbf{k}_2) \mathcal{P}_{ij, kl} (\textbf{k})
\end{equation}
where we defined the last term on RHS as $\mathcal{P}_{ij, kl} (\textbf{k}) = |h_{\textbf{k}}|^2 \sigma_{ij, kl}(\textbf{k})$, where the $\sigma$ tensor encodes all the information about the polarization tensors:
\begin{equation}
    \sigma_{ij, kl}(\textbf{k}) = \sum_{s} e^{(s)}_{ij}e^{*(s)}_{kl}
\end{equation}
and the bispectrum as:
\begin{equation}
    \langle h_{i_1 j_1} (\mathbf{k}_1) h_{i_2 j_2}(\mathbf{k}_2) h_{i_3j_3}(\mathbf{k}_3) \rangle = (2 \pi)^7 \delta^3 (\textbf{k}_1 + \textbf{k}_2+\textbf{k}_3) \mathcal{P}^2_t \tilde{\mathcal{A}}_{i_1 j_1 i_2 j_2 i_3 j_3}
\end{equation}
where $\mathcal{P}^2_t$ is the amplitude of the standard tensor power spectrum  i.e. $\mathcal{P}_t =  H^2/(2 \pi M_{Pl})^2$ and the term $\tilde{\mathcal{A}}_{i_1 j_1 i_2 j_2 i_3 j_3}$ can be decomposed into two parts: 
\begin{equation}
    \tilde{\mathcal{A}}_{i_1 j_1 i_2 j_2 i_3 j_3} = \mathcal{A}\times P_{i_1 j_1 i_2 j_2 i_3 j_3}\; .
\end{equation}
The tensorial part of the last term is given by: 
\begin{equation}
    P_{i_1 j_1 i_2 j_2 i_3 j_3} = \sigma_{i_1 j_1, kl} (\textbf{k})\sigma_{ij, kl} (\textbf{k}) \Big[ k_{3k} k_{3l} \sigma_{ij, kl} (\textbf{k}_3) - \frac{1}{2} k_{3i}k_{3k} \sigma_{ij, kl} (\textbf{k}_3) \Big] + \rm{perms. of } \{1, 2, 3\}
\end{equation}
and so, overall, it depends upon the specific polarization state $(s_1, s_2, s_3)$ we chose for the basis, i.e. $ P_{i_1 j_1 i_2 j_2 i_3 j_3} \equiv P^{s_1 s_2 s_3}$. To make contact with the literature, we will always make use of the (+++) polarization.
The term $\mathcal{A}$ is instead referred to as the scalar counterpart which can be integrated out from the tensorial factor and can be computed by means of the in-in formula we showed above. 
Then, taking just the single polarization +++, and defining $K_T = k_1 + k_2 + k_3$ we obtain: 
 \begin{equation}
   P^{+++} =  \frac{K^5_T}{128 (k_1 k_2 k_3)^2} \Big[ K^3_T - \sum_{i \neq j}k^2_i k_j - 4k_1 k_2 k_3\Big] \; .
 \end{equation}
From now on, once we set a given tensor polarization state (i.e. the $P^{s_1 s_2 s_3}$ function ) we will just have to compute the $\mathcal{A}$ factor.

\begin{figure}[h!]
    \centering
    \includegraphics[scale=0.6]{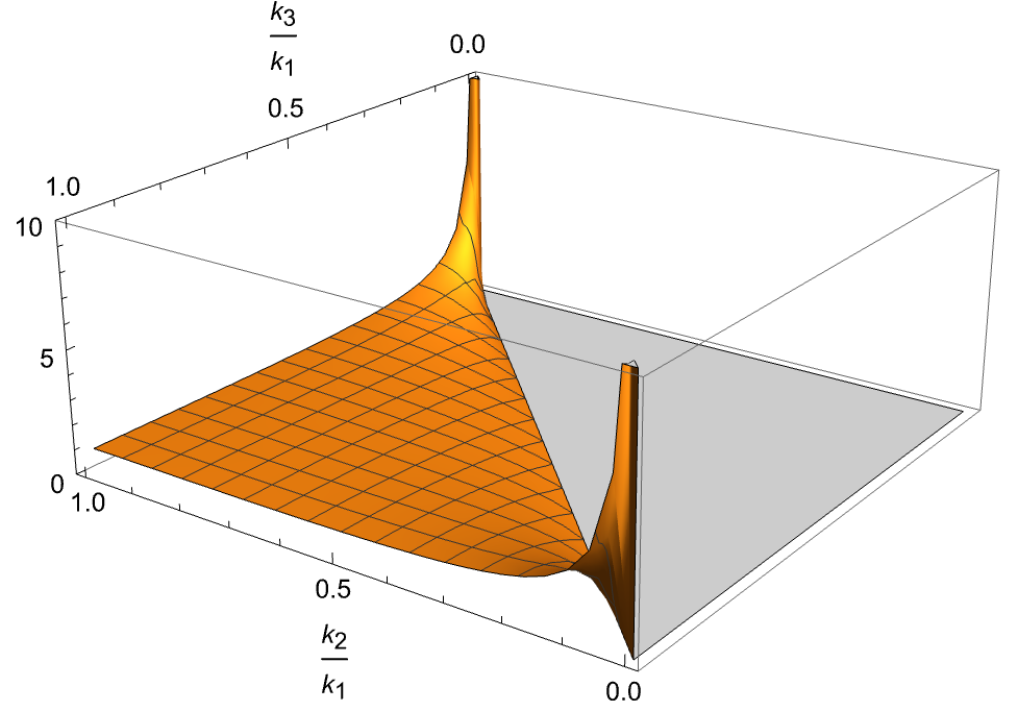}
    \caption{ Plot of the shape function $\mathcal{A}$ Eq. \ref{bispettroBD} multiplied by $(k_2/k_1)(k_3/k_1)$ to match the other results in literature \cite{Gao:2011vs}. The plot is normalized to unity for equilateral configuration.} 
    \label{fig:bispettroBD}
\end{figure}

From Eqs.~(\ref{ininformula}),(\ref{Hint}) one readily derives 
\begin{equation}
_{in}\langle 0| \hat{h}^{s_1}_{\mathbf{k}_1}(\tau) \hat{h}^{s_2}_{\mathbf{k}_2}(\tau) \hat{h}^{s_3}_{\mathbf{k}_3}(\tau) |0\rangle_{in} = \lim_{k \tau \rightarrow 0} \frac{1}{M^4_{Pl}}. \Im \Big[ h_{\textbf{k}_1}(\tau)h_{\textbf{k}_2}(\tau)h_{\textbf{k}_3}(\tau) \int^{\tau}_{-\infty} d\tau' h_{\textbf{k}_1}(\tau ')h_{\textbf{k}_2}(\tau ')h_{\textbf{k}_3}(\tau ') \Big]
\end{equation}
 where, in the last formula, we have temporarily re-instated the Planck mass in order to make an easy comparison with the literature. 
Implementing the Bunch-Davies condition for the vacuum is tantamount to requiring the following behaviour deep inside the horizon,
\begin{equation}
    v_{k\gg a H} \rightarrow \frac{e^{-ik \tau}}{\sqrt{2k}} \; .
\end{equation}
This prescription is equivalent to asking the mode functions to be in their flat (Minkowskian) configuration at the beginning of their evolution. As we shall see shortly, this corresponds to the ``minimal'' choice for the Bogoliubov coefficients $A_k$ and $B_k$ (i.e. $A_k = 1$ and $B_k = 0$).  
The choice of the BD vacuum is further motivated by the fact it is a quantum attractor for the system \cite{Kaloper:2018zgi}.
From the previous expression, one arrives at the following value for the amplitude of tensor non-Gaussianities, 
\begin{equation}
    \mathcal{A}_{BD} = -\Big(\frac{H}{M_{Pl}}\Big)^2 \frac{\Big[ \sum_{i}k^3_i + 2\sum_{ i\neq j} k^2_i k_j \Big]}{8K^2_T (k_1 k_2 k_3)^3}\ \; ,
\end{equation}
where the subscript ``BD'' underscores the current choice of the vacuum, in contradistinction to the options we will explore in \textit{Section} \ref{alphav}.  The momenta dependence of the bispectrum, that is, the shape of non-Gaussianity, is illustrated in Fig.~\ref{fig:bispettroBD}, which clearly shows a profile that is very well described by the local template \cite{Babich:2004gb}. It is convenient at this stage to define the non-linear parameter  $f_{\rm NL}$ for the tensor three-point function. Following standard practice \cite{Kanno:2022mkx, Chen:2006nt}, we define it as: 
\begin{equation}
    f_{\rm NL} = \frac{ \tilde{\mathcal{A}}_{+++}}{P_t(k_1)P_t(k_2) +{\rm perm.}} \; ,
    \label{fNLstandard}
\end{equation}
where the numerator represents the bispectrum :
    \begin{align}
    \tilde{\mathcal{A}}_{+++} & = \frac{\left(k_1^3-\left(k_2+k_3\right) k_1^2-  
    \left(k_2-k_3\right){}^2 k_1+\left(k_2-k_3\right){}^2
    \left(k_2+k_3\right)\right) }{1024 k_1^5 k_2^5 k_3^5} \times \\
    & \times \left(k_1^3+2 \left(k_2+k_3\right) k_1^2+2 \left(k_2^2+k_3
    k_2+k_3^2\right) 
    k_1+k_2^3+k_3^3+2 k_2 k_3^2+2 k_2^2 k_3 \right) \; . 
    \label{bispettroBD}
   \end{align} 
We stress here that the definition of $f_{\rm NL}$ is not univocal as one may simply build the estimator associated with any given definition. Depending on the shape of the bispectrum, one may define different non-linear parameters. One typically tries to capture, with a given $f_{\rm NL}$ definition, a finite well-defined amplitude that best signals the departure from Gaussianity. In what follows we shall also make use of a complementary definition to the one given in Eq.~(\ref{fNLstandard}), namely 
\begin{equation}
    f^{\textit{fold}}_{\rm NL} = \frac{\tilde{\mathcal{A}}_{+++}}{(P_t(k_1)P_t(k_2) + \textit{perm.}) + 3 (P_t(k_1)P_t(k_2)P_t(k_3))^{2/3} - [(P_t(k_1)P^2_t(k_2)P^3_t(k_3))^{1/3} + \textit{perm.}]}
    \label{foldeddefinition}, 
\end{equation}
where the $P_t$ we used for $f_{\rm NL}$ is the dimensional definition for the power spectrum, i.e.  $k^3P_t = \mathcal{P}_t$.
The quantity $f^{\textit{fold}}_{\rm NL}$ is indeed, particularly useful whenever the bispectrum profile is of the folded type, i.e. when the maxima of the shape-function correspond to a whole line in the $(k_2/k_1, k_3/k_1)$ plane.
 
To provide an explicit example, in the case of Bunch-Davies vacuum, for Eq.~(\ref{fNLstandard}) one immediately finds

\begin{align*}
     f^{BD}_{\rm NL} =& \frac{K_T^3 \left(2 k_2-K_T\right) \left(2 k_3-K_T\right)     \left(2 k_2+2 k_3-K_T\right)}{16384 k_1^2 k_2^2 k_3^2
      \left(k_1^3+k_2^3+k_3^3\right)} \times \\
    &\times\left(-k_2
     K_T^2-k_3 K_T^2+k_2^2 K_T+k_3^2 K_T+k_2 k_3^2+k_2^2 k_3+K_T^3\right)
\end{align*}
and a corresponding expression is found employing Eq.~(\ref{foldeddefinition}). 

As we can notice, the vanilla model predicts fixed values for the non-linear parameter. 
The only two configurations in which the $f_{\rm NL}$ parameter is effectively non-zero, are the equilateral and the squeezed ones (the folded limit, for the BD initial condition, is always null in either definition for the $f_{\rm NL}$ we take). However, as we shall see, the folded limit is the one which is more sensitive to a variation of the initial conditions.

We stress that yet another definition of $f^{\rm tens}_{\rm NL}$ is widely in use (see e.g. \cite{Kanno:2019gqw}):
\bea
\tilde{f}_{\rm NL} = \frac{\tilde{\mathcal{A}}_{+++}}{P_{\zeta}(k_1)P_{\zeta}(k_2) +{\rm perm.}} = \frac{\tilde{\mathcal{A}}_{+++}}{P_t(k_1)P_t(k_2) +{\rm perm.}} r^2 \equiv f_{\rm NL} r^2
    \label{fNLCMB}
\eea
and a corresponding definition is naturally in place, e.g., the folded case. In \textit{Table 1} we show the results depending upon the tensor-to-scalar ratio $r$.
\begin{table}[h!]
  \centering
  \caption{Values of $\tilde{f}_{\rm NL}$ obtained for the three different configurations and the BD vacuum. We obtained these numbers by setting $k_{\text{long}} \approx 10^{-3} k_{\text{short}}$ for the squeezed limit. }
  \label{tab:tabellaBD}
  \begin{tabular}{c|c}
    \hline
    \multicolumn{1}{c}{$\tilde{f}_{NL}$ for standard GR + BD vacuum} \\
    \hline
    Equilateral & Squeezed \\
    \hline
    $0.00933 \times r^2$ & $0.00293 \times r^2$ \\
    \hline
  \end{tabular}
\end{table}

\subsection{$\alpha$ vacua and fixed initial time choice}
\label{alphav}
Having introduced all the necessary tools in the previous section, we can now tackle generalized/mixed states in the early Universe. 
Since the inflationary dynamics take place in an approximate de Sitter background, we have a family of different, equivalent, vacua. This is due to the lack of time--like Killing vectors in a general dynamical background see e.g. \cite{Parker:2009uva}. 

The general solution of the Mukhanov-Sasaki equation can be written in terms of Hankel functions as  
\begin{equation}
    h_{k}(\tau) = \frac{A_{k}}{a(\tau)} \frac{e^{- i k \tau}}{\sqrt{2k}} \Big(1 - \frac{i}{k \tau}\Big) + \frac{B_{k}}{a(\tau)} \frac{e^{ i k \tau}}{\sqrt{2k}} \Big(1 + \frac{i}{k \tau}\Big) \equiv A_k u_{k} + B_k u^*_k,    
\end{equation}
where the two time-independent parameters $A_k$ and $B_k$ are known as Bogoliubov coefficients. These coefficients are bound to the normalization condition imposed by requiring the conservation of the canonical relations for the ladder operators $ a_{\textbf{k}} $
and $a^{\dag}_{\textbf{k}}$, that is, $[a_{\textbf{k}'} ,a^{\dag}_{\textbf{k}} ] = \delta^3(\textbf{k}'-\textbf{k})$. 
As a result, one finds: 
\begin{equation}
    |A_k|^2 - |B_k|^2 = 1\; .
\end{equation}
This relation allows for a simple parametrization in terms of hyperbolic functions, giving, for the wavefunction $h$:
\begin{equation}
    h_k (\tau) = u_k \cosh{\alpha} + u^*_k e^{i \beta} \sinh{\alpha} , 
\end{equation}
where the two parameters $\alpha,\beta$ can vary in the respective range: $\alpha \in [0, + \infty)$ and $\beta \in (- \pi, \pi)$. 

We note here that non-zero values for $\beta$ would break time-reversal invariance so that the quantum state wouldn't be invariant under the symmetry group $SO(4,1)$. Setting set $\beta = 0$ is also necessary in order to smoothly reduce to the choice of a Bunch-Davies vacuum. For a complete treatment of the properties of the vacua in quasi-de Sitter (dS) and mixed states, see \cite{Allen:1985ux} as well as \cite{Mukhanov:2007zz, Kanno:2022mkx, Bahrami:2013isa}.

At this stage one typically converges towards specific combinations of the Bogoliubov coefficients, see \cite{Mukhanov:2007zz, Chung:2003wn} for some well-motivated examples. 
In this section, we shall take the point of view of \cite{Danielsson:2002kx, Broy:2016zik} which underscores the fact that initial conditions are often imposed for very large momenta. In this regime, one does not have complete control of the quantum field theory description. It is then convenient to set a cutoff scale $\Lambda$ that limits the remit of the standard treatment. One may equivalently identify the corresponding (conformal) time $\bar{\tau}$ defined as:

\begin{equation}
    \bar{\tau}_k = -\frac{\Lambda}{kH} \; 
    \label{initialtime} 
\end{equation}

Additional related possibilities have been considered in
\cite{BouzariNezhad:2018zsi}. The 
quantization procedure is standard, with the caveat that the vacuum annihilation holds at a precise time in the evolution, 
\begin{equation}
    \hat{a}_{\textbf{k}}(\bar{\tau}_k) |0 \rangle = 0\; ,
\end{equation}
something commonly known in literature \cite{Kundu:2011sg, Danielsson:2002kx, Broy:2016zik, Mukhanov:2007zz} as \textit{instantaneous vacuum}. One quickly arrives \cite{Broy:2016zik} at the following expression for the Bogoliubov coefficients,
\begin{equation}
    A_k = \sqrt{1 + B^2_k} \qquad \qquad B_k = \frac{H}{2\Lambda}\; .
\end{equation}
Up to slow-roll corrections, both coefficients are not explicitly dependent on the scale $k$. The key quantity is the $H/\Lambda$ ratio, with $H$ a good proxy for the energy scale of inflation and $\Lambda$ signalling when quantum gravitational effects become important. The resulting two-point function is
\begin{equation}
    P^{f.t.} = P^0_t \Big[ 1 + \frac{H}{\Lambda} \sin{\frac{\Lambda}{H}} \Big]\; ,
\end{equation}
where the oscillating feature regulated by  $H/\Lambda$ is completely negligible for typical (i.e. large) values of $\Lambda$. 
We can now move on to explore the effects of a non-trivial initial state on gravitational wave non-Gaussianities. We will employ the very same framework we described in \textit{Section} \ref{alphav}  to estimate the non-linear parameter and the bispectrum shape function. 
\begin{figure}
    \centering
    \includegraphics[scale=0.6]{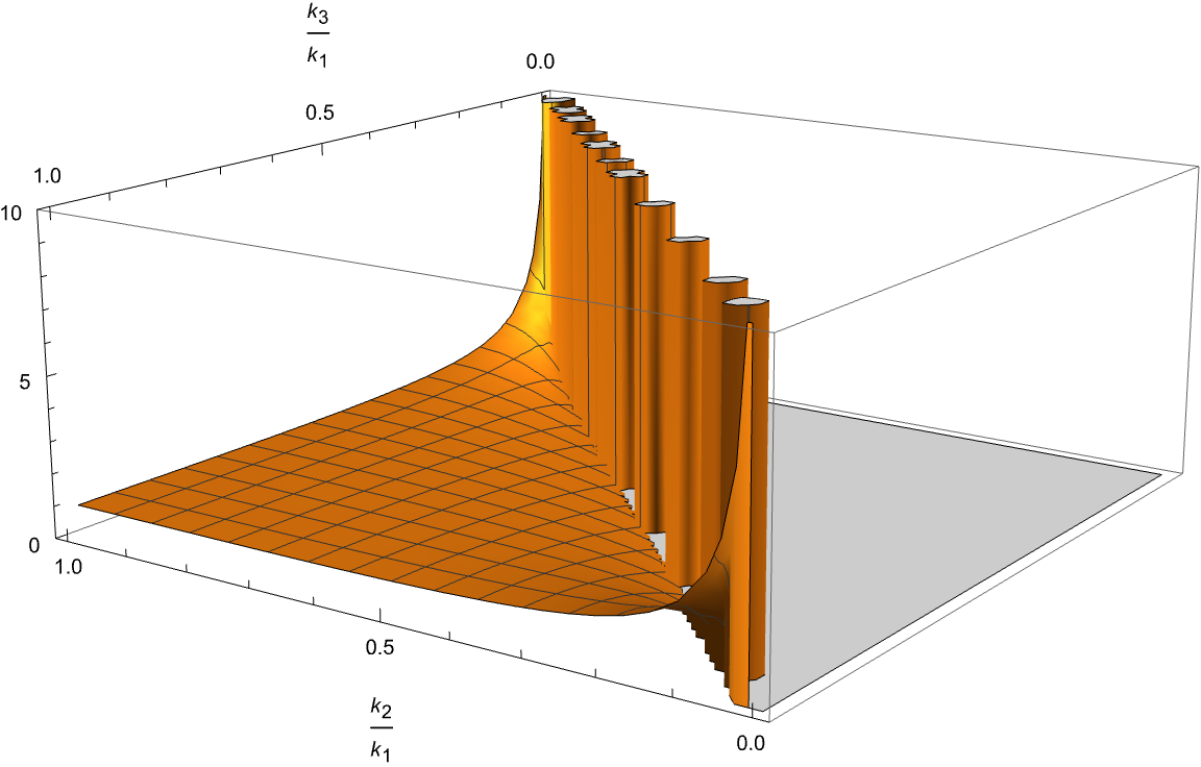}
    \caption{Plot of the shape function $\mathcal{A}$ Eq. \ref{bispettroalphavacua} multiplied by $(k_2/k_1)(k_3/k_1)$ . The plot is normalized to unity for equilateral configuration.}
    \label{fig:bispettroalphavac}
\end{figure}
The shape function in Fig.~\ref{fig:bispettroalphavac} peaks along the diagonal line of the $(k_2/k_1,k_3/k_1)$  plane. This behaviour is characteristic of the so-called folded shape and it has been shown to arise e.g. in the scalar sector in the case of non-BD vacua (\cite{Jiang:2015hfa, Wang:2013zva, Kanno:2022mkx, Ghosh:2022cny}). This contribution is a characteristic signature of the presence of a non-zero $\beta_{k}$ coefficient which is responsible for the particle number density (\cite{Boyle:2005se, Wang:2013zva}).
When dealing with this quantity it turns out to be convenient to introduce a suitable momentum cut-off $\omega$ and corresponding  regularized momentum $\textbf{k}_1 = \textbf{k}_2 + \textbf{k}_3 + \omega$, following a procedure that removes any unphysical divergence \cite{Wang:2013zva}. 
Our explicit computation arrives at the same momenta behaviour obtained in \cite{Kanno:2022mkx}. One difference is that, in place of the geometrical $\alpha$ parameter for the parametrization of the initial quantum state, we choose to highlight instead the precise dependence on the energy scale $\Lambda$. 
In Appendix \ref{app:a} we report the explicit expression for both the $f_{\rm NL}$ parameter and the Bispectrum shape function $ \mathcal{A}$.

\begin{figure}
    \centering
    \includegraphics[scale=0.47]{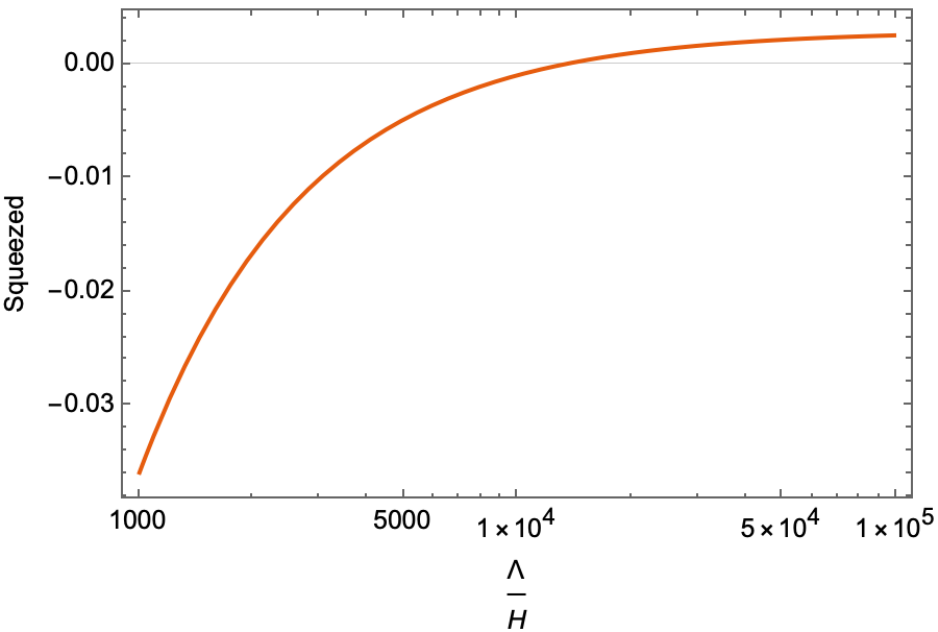}
    \includegraphics[scale=0.49]{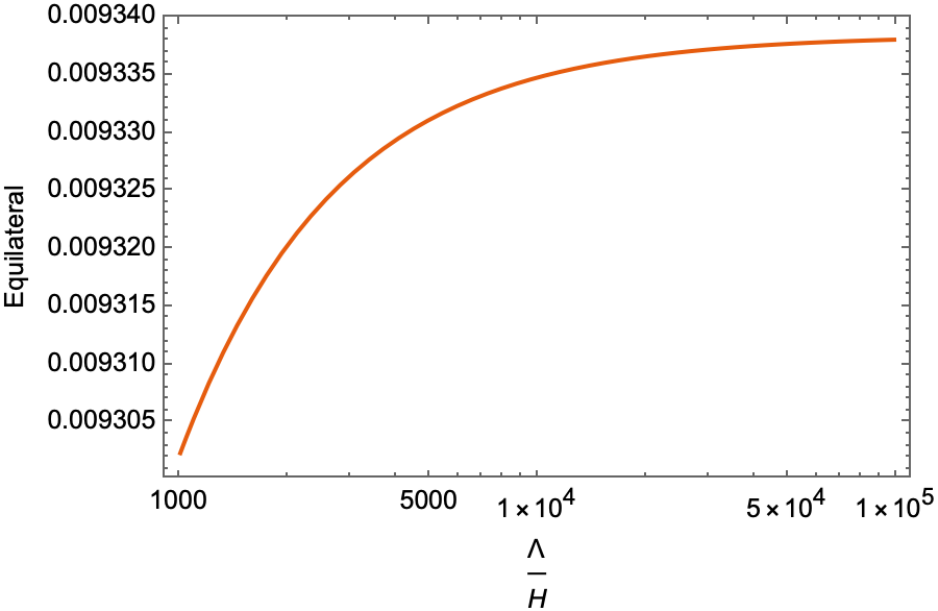}
    \caption{Evolution of the $f_{\rm NL }$ parameter as a function of the ratio $\Lambda/H$ for the equilateral and squeezed configurations. }
    \label{fig:fnlfizedtime}
\end{figure}

Inspection of Eq.~(\ref{fnlfixedtime}) reveals the expected dependence of the non-linear parameter on the $\Lambda/H$ ratio. In Fig. \ref{fig:fnlfizedtime} we plot the ``running'' of different configurations as a function of this ratio with a cut-off energy limited by the Planck mass and the Hubble rate limited to be at most $10^{13} {\rm GeV}$. The non-linear parameter may acquire negative values in the squeezed configuration. In every configuration, when $\Lambda$  approaches the Planck mass the non-linear parameter values approach the predictions obtained in the standard case.

\bigskip

\section{Stochastic source}
\label{source}
In this section, we shall model the physics that may result, for example, from the presence of additional degrees of freedom sourcing the tensor sector. The anisotropic stress tensor in the GW equation encodes such contributions. Negligible in the standard single-field slow-roll scenario, these may play a leading role in a host of (pre-)inflationary realizations. 
\bigskip

One may start from the notion that modes created at very early times have necessarily experienced a trans-Planckian regime \cite{Alberghi:2003am, Brandenberger:2012aj, Tanaka:2000jw, Danielsson:2002kx} on which we have limited access.
 
Following the approach of \cite{Cielo:2022vmo} we will account for the physics stemming from a trans-Planckian regime through the use of a stochastic source which, at the Lagrangian level, is formally an interaction between the tensor field $h_{ij}$ and an effective operator $\Pi_{ij}$.  The main result of the work in \cite{Cielo:2022vmo}  is that the prediction for the amplitude of the gravitational wave power spectrum from the BD vacuum is just a particular case in a continuous spectrum of possibilities depending on the $\Lambda/H$ ratio.
Much as in \cite{Cielo:2022vmo}, our motivations are the following. The description of primordial quantum perturbations cannot be treated as a closed quantum system, but rather an open one. This means that the linear approximation for the perturbations is not adequate, especially when the physical modes approach a given cut-off scale, close to a pure quantum gravitational regime, i.e. $k/a(\tau) \approx \Lambda$. One can consider such non-linearities, in manifold ways. First, one can naturally take tackle higher order perturbation theory.
An alternative approach is to describe non-linearities by means of collective forces, i.e. adopt the the mean-field treatment. The latter one will be our approach: it consists in a parametrization of the source term $\Pi_{ij}$ as a Gaussian, stochastic, time - uncorrelated, term with possibly an explicit $k$ dependence. This term ought to switch off at the time $\bar{\tau}_k$, when perturbations become sub-Planckian and both GR and the linear regime can be trusted:

\begin{align}
h''_k + 2 \mathcal{H} h'_k + k^2 h_k = 16\pi G a^2 \, \Pi_k ~~~&& \tau < \bar{\tau}_k \label{e:source} \\
h''_k + 2 \mathcal{H} h'_k + k^2 h_k = 0 ~~~~~~~~~~~~~&& \tau > \bar{\tau}_k. \label{e:nosource} 
\end{align}
The source term is treated as a  quantum object much as are tensor fluctuations. As standard, we introduce the ladder operators 
\begin{equation}
    \hat{\Pi}^r_{\textbf{k}} (\tau) = \Pi_k(\tau) \hat{a}^r_{\textbf{k}} + \Pi^*_k(\tau) \hat{a}^{r \dag}_{-\textbf{k}}\; .
\end{equation}
Here too, we follow closely the same approach of \cite{Cielo:2022vmo} where the sea from which the trans-Planckian modes emerge is realized as a Brownian environment, and the source is fully specified by the following conditions:
\begin{align}
    \langle \Pi_k(\tau) \rangle &= 0  \\
    \langle \Pi_k (\tau)\Pi_k^* (\tau') \rangle &= \Lambda^6 F(k) \delta (\tau - \tau') \; ,
\end{align}
where we have introduced a dimensionful normalization constant in the two-point function for the noise. Note that, in the current context the brackets have a double role: the usual expectation value over quantum state configurations and the statistical ensemble average over stochastic configurations of the noise. 
Here we will limit the analysis to a white noise spectrum  $F(k) = 1$. Further analysis considering the {\it black hole plasma} model described in \cite{Cielo:2022vmo} will be considered elsewhere.

The global solution for $h$ is then obtained by implementing standard matching conditions:
\begin{align}
 \lim_{\tau \rightarrow \bar{\tau_k}^-} h_k (\tau) &= \lim_{\tau \rightarrow \bar{\tau_k}^+}h_k (\tau) \label{matching0}\\
\lim_{\tau \rightarrow \bar{\tau_k}^-} h'_k(\tau) &= \lim_{\tau \rightarrow \bar{\tau_k}^+} h'_k(\tau). \label{matching1} 
\end{align}
The general solution of the non-homogeneous Eq.~(\ref{e:source}) is given by:
\begin{align}
h_k (\tau) = \frac{16\pi G}{a(\tau)}\,\int^{\tau}_{-\infty} d \tau' a(\tau ') G_k(\tau, \tau') \Pi_k (\tau'),
\end{align}
where the Green's function $G_k$ is  (see e.g. \cite{Biagetti:2013kwa}): 
\begin{equation}
    G_{k}(\tau, \tau') = \frac{e^{- i k(\tau + \tau')}}{2 k^3 {\tau'}^2} \Big[e^{2ik \tau}(1 - i k \tau)(-i + k \tau')  + e^{2 i k \tau'}(1 + i k \tau)(i + k\tau') \Big] \Theta(\tau - \tau') . 
\end{equation}
The matching of the solution across $\bar{\tau}$ enforces specific conditions on the Bogoliubov coefficients $A_k, B_k$. 
Interestingly, from this point of view a non-homogeneous equation in the trans-Planckian regime is equivalent to the selection of a non-Bunch-Davies initial state, see \cite{Cielo:2022vmo} for a thorough study of such equivalence.

\bigskip

\begin{figure}
    \centering
    \includegraphics[scale=0.6]{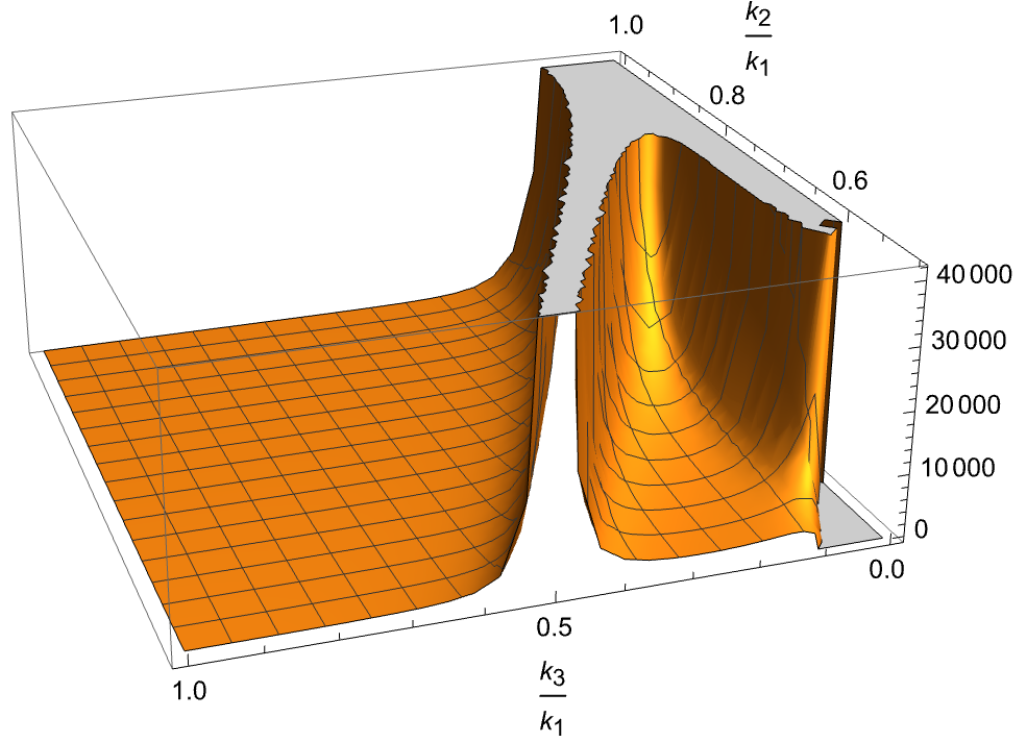}
    \caption{Characteristic shape of the bispectrum in the presence of a stochastic source. 
    }
    \label{fig:bispettrosourced}
\end{figure}

\begin{figure}
    \centering
    \includegraphics[scale=0.6]{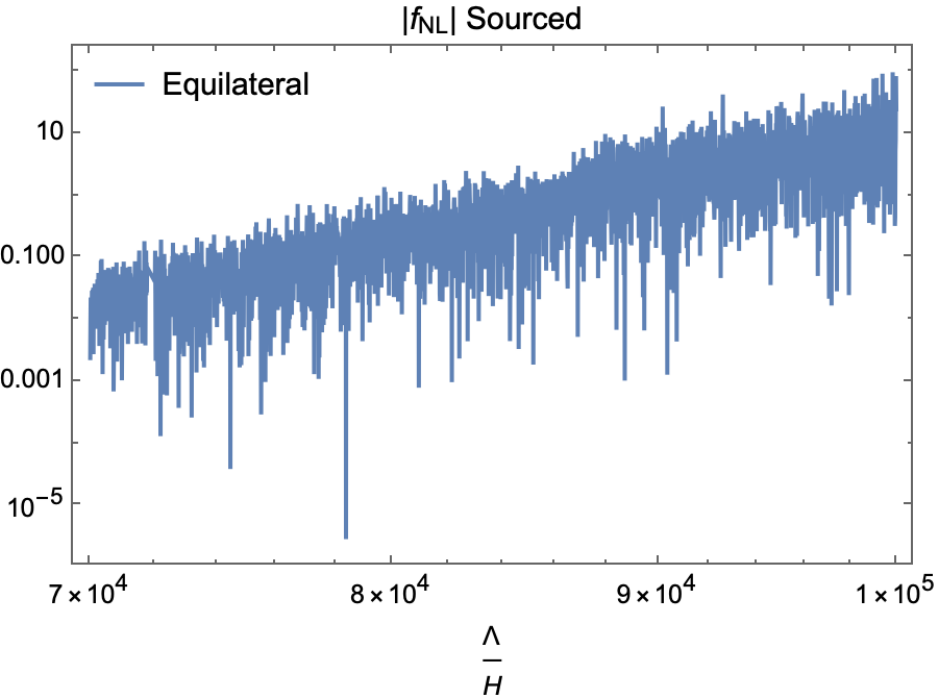}
    \caption{LogPlot of the non-linear parameter for the equilateral configuration.}
    \label{fig:squeezedsourced}
\end{figure}

\begin{figure}
    \centering
    \includegraphics[scale=0.6]{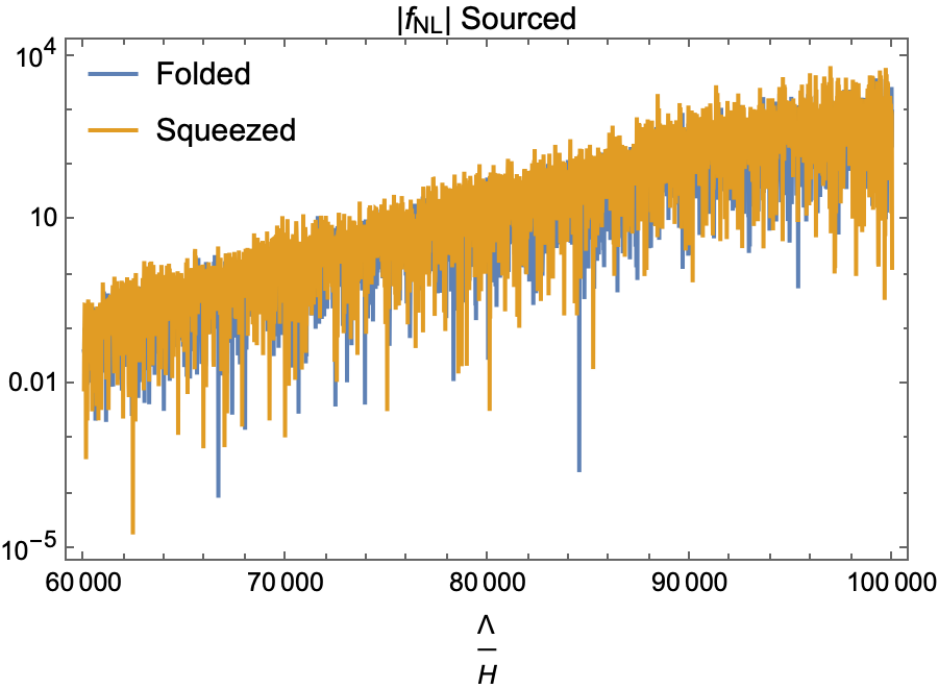}
    \caption{
    LogPlot of the $|f_{\rm NL}|$ parameter for the squeezed and folded configurations. }
    \label{fig:equilateralandfolded}
\end{figure}

As one can see in Fig.~\ref{fig:squeezedsourced} and Fig.~\ref{fig:equilateralandfolded} the most significant non-linear parameters are those in the squeezed and the folded limits\footnote{Where the latter has been derived according to Eq.~(\ref{foldeddefinition}).}.
\bigskip

Finally, we can conclude that this class of models can support large tensor non-Gaussianities. 

\section{Conclusions}
The standard treatment of cosmological observables from the very early universe is often oblivious to pre-inflationary dynamics.  This is not surprising given that, despite great progress e.g. in string theory, a predictive and testable UV complete theory for gravitational interactions remains elusive. In order to try and include pre-inflationary physics in our description, one may codify the effect of a trans-Planckian era by means of a quantum noise component.
In this work, we explored the effects on cosmological correlators due to the presence of such a stochastic source. Specifically, we studied the amplitude and shape of the primordial tensor three-point function.

A signal characterization beyond the power spectrum is crucial in light of the numerous inflationary models, as well as initial conditions, that correspond to the same prediction at the level of the two-point function. We found that both the amplitude and shape function of primordial gravitational wave non-Gaussianities are rather sensitive to the stochastic source. Depending on $\Lambda$, the scale regulating the onset of trans-Planckian effects, a suitably defined non-linear parameter $f_{\rm NL}$ can be as large as $\mathcal{O}(100)$ in the equilateral limit and $\mathcal{O}(10^{4})$ both in the squeezed and folded configurations.  These numbers are several orders of magnitude larger than the standard single-field slow-roll result and are within reach for upcoming gravitational wave probes. 

Our focus has been on the case of a white noise spectrum. It would be intriguing to consider a scale-dependent $F$ function and its effect on the observables at hand. This is particularly relevant in view of the fact that it is difficult to directly test gravitational wave non-Gaussianity at intermediate and small scales \cite{Bartolo:2018evs}. This is due to the fact that the gravitational wave modes take a different path through structure and, as a result, the initial primordial correlation is highly suppressed by the time the modes reach a detector. Remarkably, this suppression is not in place for the ultra-squeezed configuration with important consequences also for gravitational wave anisotropies. 

Further efforts towards including stochastic effects in the study of primordial gravitational waves higher-point functions will be indispensable to unveil the dynamics underlying the (pre-)inflationary stage. We plan to expand on the present analysis in a forthcoming work.

\acknowledgments
MC thanks the IFT Madrid for its hospitality and support. MC, GM, and OP are partly supported by the Italian Ministero dell’Universit\`a e Ricerca (MUR) through the research grant number 2017W4HA7S “NAT-NET: Neutrino and Astroparticle Theory Network” under the program PRIN 2017, and by the Istituto Nazionale di Fisica Nucleare (INFN) through the “Theoretical Astroparticle Physics” (TAsP) project.  MF acknowledges support from the “Ram\'{o}n y Cajal” grant RYC2021-033786-I. MF’s work is partially supported by the Spanish Research Agency (Agencia Estatal de Investigaci\'{o}n) through the Grant IFT Centro de
Excelencia Severo Ochoa No CEX2020-001007-S, funded by MCIN/AEI/10.13039/501100011033.

\appendix

\section{Results for $\alpha$ -vacua}
\label{app:a}
In this Appendix, we will list the two main formulas obtained at the end of \textit{Section} 2.
Here we display the result for the amplitude for $\tilde{\mathcal{A}}$ computed by choosing the initial condition at a fixed initial time, Eq. (\ref{initialtime}). 
\begin{equation}
    \begin{aligned}
    \tilde{\mathcal{A}}^{f.t.}_{+++} &= \Bigg[ \left(K_T\right)^3 \left(\left(K_T\right)^3-4 k_1 k_2 k_3-4 \left(k_2 k_1^2+k_3 k_1^2+k_2^2 k_1+k_3^2 k_1+k_2 k_3^2+k_2^2 k_3\right)\right) \\
    & \times \left(-\left(H-\Lambda \sqrt{\frac{H^2}{\Lambda ^2}+4}\right)^3\right) \\
    & \times \left(16 H^3 k_1^3 k_2^3 k_3^3+16 H^2 k_1^3 k_2^3 k_3^3 \Lambda \sqrt{\frac{H^2}{\Lambda ^2}+4} \right. \\
    & \left. + \left(k_1^3-\left(k_2+k_3\right) k_1^2-\left(k_2-k_3\right)^2 k_1+\left(k_2-k_3\right)^2 \left(k_2+k_3\right)\right)^2 \right. \\
    & \times \left(k_1^3+2 \left(k_2+k_3\right) k_1^2+2 \left(k_2^2+k_3 k_2+k_3^2\right) k_1+k_2^3+k_3^3+2 k_2 k_3^2+2 k_2^2 k_3\right) \Lambda ^3 \\
    & \left. \times \left(-\sqrt{\frac{H^2}{\Lambda ^2}+4}\right) + H \left(k_1+k_2+k_3\right)^2 \right. \\
    & \times \left(k_1^7-2 \left(k_2+k_3\right) k_1^6+6 k_2 k_3 k_1^5+\left(k_2^3-6 k_3 k_2^2-6 k_3^2 k_2+k_3^3\right) k_1^4 \right. \\
    & + \left(k_2+k_3\right)^4 k_1^3-6 k_2 \left(k_2-k_3\right)^2 k_3 \left(k_2+k_3\right) k_1^2 \\
    & -2 \left(k_2-k_3\right)^4 \left(k_2^2+k_3 k_2+k_3^2\right) k_1+\left(k_2-k_3\right)^4 \left(k_2^3+2 k_3 k_2^2+2 k_3^2 k_2+k_3^3\right) \Lambda ^2\Bigg] \\
    & \times \Bigg[ 16384 \Lambda ^6 k_1^5 k_2^5 k_3^5 (k_1+k_2-k_3)^2
    (k_1-k_2+k_3)^2 (-k_1+k_2+k_3)^2 \Bigg]^{-1}
    \end{aligned}
    \label{bispettroalphavacua}
\end{equation} 
Then, the following represents the relative ${f}_{\rm NL}$ parameter calculated according to Eq. (\ref{fNLstandard}): 
\begin{equation}
    \begin{aligned}
    f^{f.t.}_{\rm NL} &= \Bigg[ -\left(H-\Lambda \sqrt{\frac{H^2}{\Lambda ^2}+4}\right)^3 K_T^3 \left(2 k_2-K_T\right) \left(2 k_2+2 k_3-K_T\right) \\
    & \times \left(16 H^3 k_1^3 k_2^3 k_3^3+16 H^2 k_1^3 k_2^3 k_3^3 \Lambda \sqrt{\frac{H^2}{\Lambda ^2}+4}+\Lambda ^3 \left(-\sqrt{\frac{H^2}{\Lambda ^2}+4}\right) \right. \\
    & \times \left(K_T-2 k_2\right)^2 \left(K_T-2 k_3\right)^2 \left(-2 k_2-2 k_3+K_T\right)^2 \\
    & \times \left(k_2^2 \left(k_3+K_T\right)+k_2 \left(k_3^2-K_T^2\right)+K_T \left(-k_3 K_T+k_3^2+K_T^2 \right) \right) \\
    & + H \left(k_1^7-2 \left(k_2+k_3\right) k_1^6+6 k_2 k_3 k_1^5+\left(k_2^3-6 k_3 k_2^2-6 k_3^2 k_2+k_3^3\right) k_1^4 \right. \\
    & + \left(k_2+k_3\right)^4 k_1^3-6 k_2 \left(k_2-k_3\right)^2 k_3 \left(k_2+k_3\right) k_1^2 \\
    & -2 \left(k_2-k_3\right)^4 \left(k_2^2+k_3 k_2+k_3^2\right) k_1 +\left(k_2-k_3\right)^4 \left(k_2^3+2 k_3 k_2^2+2 k_3^2 k_2+k_3^3\right) \\
    & \left. \times \Lambda ^2 K_T^2 \right] \times \\
    & \times \Bigg[  262144 k_1^2 k_2^2 k_3^2 \left(-k_1+k_2+k_3\right)^2 \left(k_1^3+k_2^3+k_3^3\right) \Lambda ^6
   \left(K_T-2 k_2\right){}^2 \left(K_T-2 k_3\right)\Bigg]^{-1}
    \end{aligned}
    \label{fnlfixedtime}
\end{equation}

\bibliographystyle{unsrt}
\bibliography{ref.bib}

\end{document}